\documentclass[12pt,letterpaper]{article}

\usepackage{amsmath}
\usepackage{amsfonts}
\usepackage{amssymb}
\newcommand{\tri}{\hspace{-3.5pt}\vartriangle\hspace{-3.5pt}}

\begin{document}

\vspace*{-1.5cm}
\begin{flushright}
  {\small
  MPP-2011-152\\
  }
\end{flushright}

\vspace{1.5cm}
\begin{center}
  {\LARGE
Nonassociativity  in String Theory}
\end{center}

\vspace{0.75cm}
\begin{center}
  Ralph Blumenhagen
\end{center}

\vspace{0.1cm}
\begin{center} 
\emph{Max-Planck-Institut f\"ur Physik (Werner-Heisenberg-Institut), \\ 
   F\"ohringer Ring 6,  80805 M\"unchen, Germany } \\[0.1cm] 
\vspace{0.2cm}

 \vspace{0.5cm} 
\end{center} 

\vspace{1cm}


\begin{abstract}
I summarize some of the ideas and motivations 
behind a recently performed conformal field theory 
analysis  of closed strings in both geometric and nongeometric three-form
flux backgrounds.   
This suggests  an underlying
nonassociative structure for the coordinates.
\end{abstract}

\footnotetext[1]{This article summarizes a blackboard talk given at the
``Memorial Conference for Maximilian Kreuzer''  held in Vienna, June 2011.}
\clearpage

\section{Introduction}\label{ra_sec1}

Max will remain in our memory for his contributions to the understanding of 
Calabi-Yau manifolds.
The classification of reflexive polytopes in the framework of toric geometry
is surely  his most acknowledged  contribution. However, his interests
were much broader  and together with M. Herbst and A. Kling
he also wrote a couple of, I think, very nice papers on noncommutative
geometry\cite{hep-th/0106159,hep-th/0203077}. 
In particular, they were analyzing open strings in the background
of a non-constant two-form  background, i.e. one with non-trivial
three-form flux,  and found that
in this case the coordinates are not only noncommutative but
also nonassociative. Mathematically, an important role in their ana\-lysis was 
played by the Rogers dilogarithm.
Intrigued by their results and similar ones by Cornalba/Schiappa 
\cite{hep-th/0101219}, my collaborators and myself
wondered whether a non-trivial three-form
flux background might also have similar  effects on the closed string 
sector, which is the one governing gravity. 

Let me  pose three questions, which I believe have at least the potential
to   point to such new structures
in closed string theory.
It is expected that  the answers to these three questions
are related, and some first concrete computations make it conceivable  
that string theory at small scales is dual
to a theory which involves nonassociative spaces, for which
the Kalb-Ramond field has been traded for a nonassociatve deformation
of ordinary Riemannian geometry. 

\subsubsection*{Closed string generalization of noncommutative geometry}

It is a well established fact that the effective theory on a D-brane
equipped with a non-trivial two-form magnetic flux ${\mathcal F}=B+F$
becomes noncommutative. This can be deduced by studying the conformal
field theory on a flat D-brane with a constant magnetic field.
In this case, the
two-point function of two open string coordinates $X^a(z)$ inserted on the
boundary of a disk  takes the form 
\begin{equation}
  \label{twopointa}
   \bigl\langle X^a (\tau_1)\, X^b(\tau_2) \bigr\rangle =  
  - \alpha' G^{ab} \log (\tau_1-\tau_2)^2 + 
   i \, \theta^{ab}\,  \epsilon(\tau_1-\tau_2)\, ,
\end{equation}
where $\tau$ stands for the real part of the complex world-sheet coordinate
$z$. The matrix $G^{ab}$ is symmetric and can be interpreted as the (inverse
of the) effective metric seen by the open string. $\theta^{ab}$ is
related to the two-form flux as $\theta^{ab}\simeq {\mathcal F^{ab}\over 1+ {\mathcal F}^2}$ and thus  is anti-symmetric.
The reason for the appearance of noncommutativity is the second
term in \eqref{twopointa} which means that the flux distinguishes
between the order of the two-points on the boundary of the disk.

This has been made more precise by analyzing open string
scattering amplitudes for open string vertex operators
\begin{equation}
     V=F(\partial X^\mu) e^{ip X}\, ,
\end{equation}
where $F$ is a function of $\partial X$.
Since the second term in \eqref{twopointa} is locally constant, 
it only contributes to correlation functions involving
the $\exp( ip X)$ factor in the vertex operators.
Its effect is that it introduces non-trivial momentum dependent
phases, which can be described by the introduction of
a noncommutative product on the space of functions
\begin{eqnarray}
  \label{Nbracketcon}
  && f_1(x)\, \star\,  f_2(x) =
  \exp\left( i  \theta^{ab}\,
      \partial^{x_1}_{a}\,\partial^{x_2}_{b}  \right)\, f_1(x_1)\, f_2(x_2)
    \Bigr|_{x_1= x_2=x} \;.
\end{eqnarray}
This is the Moyal-Weyl product, which implies $[x_a,x_b]=x_a\star x_b-
x_b\star x_b=i\theta_{ab}$.

Thus, noncommutavity arises for open strings in a magnetic flux background
leading to noncommutative gauge theories. One might have expected
that noncommutative geometry should also play an important role
for quantum gravity, but for closed strings a similar structure
has not been identified. 
Thinking about this question, one realizes that 
the closed string analogue must  clearly be different as
here two vertex operators are inserted in the bulk of 
a two-sphere $S^2$ and no unambiguous ordering can be defined.
Therefore, one does not expect the same kind of noncommutativity to arise.
Moreover, for a closed string a constant $B$-field can  be gauged away.

However, if  one considers  three nearby points  on the world-sheet
$S^2$ of a closed string, one can 
very well decide whether the loop connecting the three points has positive
or negative orientation. Thus, if there exists a background field
which distinguishes these two orientations, one would expect a non-vanishing
result not for the simple commutator, but  for
the cyclic double commutator
\begin{equation}
\label{jacobiid}
\begin{split}
  \bigl[X^\mu,X^\nu,X^\rho\bigr]:=
  \lim_{\sigma_i\to \sigma}\; \bigl[ [X^\mu (\sigma_1,\tau),
    X^\nu(\sigma_2,\tau)],X^\rho(\sigma_3,\tau)\bigr] +{\rm cyclic}\;. 
\end{split}
\end{equation}
Now, the
question is whether there exists a three-form with this property?

\subsubsection*{Nonlinear sigma models}

The usual approach to string theory is perturbative, i.e.
one considers a string moving in a background with metric 
$G_{\mu\nu}$, Kalb-Ramond field $B_{\mu\nu}$ and dilaton $\Phi$, whose
dynamics is governed by a two-dimensional non-linear sigma model.
With $\Sigma$ denoting the world-sheet of the closed string, its action reads
\begin{equation}
  \label{action_730178}
  \mathcal S=\frac{1}{2\pi\alpha'}\int_\Sigma d^2 z\, \bigl(\,
  G_{ab} + B_{ab} \bigr)\, \partial X^a \,\overline\partial X^b + \ldots
  \;,
\end{equation}
where we suppressed  the dilaton part.
This is treated perturbatively in a dimensionless
coupling $\sqrt{\alpha'}/R$, where $R$ is a characteristic length scale
of the background. The guiding principle is conformal invariance.
This means that the 
string equations of motion for the space-time fields
$G_{\mu\nu}$, $B_{\mu\nu}$ and $\Phi$  are given by the vanishing beta-function
equations. At leading order these equations read
\begin{eqnarray}
  \label{targeteom}
 0 &=&\beta_{ab}^G  =  \alpha' \Bigl( R_{ab}-\frac{1}{4}\: H_{a}{}^{cd}\, H_{bcd}  
  +2 \nabla_a\nabla_b\Phi\Bigr)
    +O({\alpha'}^2)\nonumber\\
 0 &=&\beta_{ab}^B  =  \alpha'\Bigl( -{1\over 2}  \nabla_c H^c{}_{ab} +
   \alpha' H_{ab}{}^c \nabla_c \Phi \Bigr)
    +O({\alpha'}^2)\\
   0 &=&\beta_{ab}^\Phi  =  {1\over 4} (d-d_{\rm crit})+ \alpha' \Bigl(
    (\nabla \Phi)^2 -{1\over 2} \nabla^2 \Phi -{1\over 24} H^2 \Bigl)
    +O({\alpha'}^2)\, .\nonumber
\end{eqnarray}
The first equation, for instance, is nothing else than 
Einstein's equation with sources.
Clearly, in this approach one is assuming from the very beginning that
the string is moving through a Riemannian geometry with additional smooth
fields. However, it is well known that there exist conformal
field theories which cannot be identified with such simple
geometries. These are  left-right asymmetric like for instance
asymmetric orbifolds. The latter are asymmetric at some
orbifold fixed points but one can imagine asymmetric CFTs which
are not even locally geometric. What
is the target space interpretation of such asymmetric CFTs?

In the non-linear sigma model one performs perturbation theory around
the  large volume limit with diluted fluxes. Can one also define
a perturbation theory around the other limit, namely very small
substringy sizes of the background?  In view of double field theory, 
we have here in mind an effective field theory describing the dynamics
of winding states in a  $G_{\mu\nu}$, $B_{\mu\nu}$,
$\Phi$ background.

\subsubsection*{What is $R$-flux?}

In the past, applying T-duality to known configurations has 
led to  new insights into string theory, where a prominent
example is the discovery of D-branes. Applying
T-duality  to the  closed string background\cite{hep-th/0508133} given by
a flat space with constant non-vanishing 
three-form flux $H=dB$, results in a background with geometric 
flux. This so-called  twisted torus is still a conventional string background,
but a second T-duality leads to a  non-geometric flux background.
These are spaces in which  the
transition functions between two charts of a manifold are allowed
to be T-duality transformations, hence they are also called T-folds.
After formally applying a third T-duality, not along an isometry direction 
anymore,
one obtains an $R$-flux background which does not 
admit a clear target-space interpretation. It was proposed 
not to  correspond to an ordinary geometry even locally,
but instead to give rise to a nonassociative 
geometry\cite{hep-th/0412092}.
In addition to involving a T-duality in a non-isotropic direction, 
another  problem with this argument is that flat space with constant $H$-flux
is not an exact  solution to the string equations of motion. 
Therefore one should ask, whether nevertheless 
one can make this R-flux case more precise.

\section{CFT analysis of $H$-flux}

The remainder of this article  is essentially a brief version of the 
more exhaustive analysis
presented recently  \cite{arXiv:1106.0316}.
First, we note that the origin of T-duality lies in conformal field 
theory where it is nothing
else than an asymmetric reflection $(X_L,X_R)\to (X_L,-X_R)$. Therefore,
it is tempting to try to analyze $R$-flux from the CFT point of view.
In order to see what is going on, let us
first perform a perturbative analysis  of the $H$-flux case and then
apply a T-duality.
We observe that a Ricci flat metric, vanishing dilaton  and a
constant $H$-flux solves the string equation of motion up to linear
order in $H$ and arbitrary order in $\alpha'$.
Therefore, the starting point is a  flat  metric and a constant
$H$-flux  specified by 
\begin{equation}
  \label{setup_01}
   ds^2=\sum_{a=1}^N \bigl(dX^a\bigr)^2, \hspace{20pt} 
   H=  \frac{2}{{\alpha'}^2}\,  \theta_{abc}\,  dX^a\wedge dX^b\wedge dX^c\; ,
\end{equation}
where for simplicity  we focus  on  $N=3$.
The expectation is that this background  corresponds to a CFT up to linear
order in $H$. 

To proceed, we write  the action \eqref{action_730178} as the sum of a 
free part $\mathcal S_0$ and a perturbation $\mathcal S_1$. 
Choosing a  gauge   such that $B_{ab} = \frac{1}{3}  H_{abc}\, X^c$, we have
\begin{equation}
  \label{action_pertub_01}
  \mathcal S = \mathcal S_0 + \mathcal S_1 \hspace{20pt}{\rm with}\hspace{20pt}
  \mathcal S_1=\frac{1}{2\pi\alpha'} \:\frac{H_{abc}}{3} \int_\Sigma d^2 z\, X^a
  \partial X^b \,\overline\partial X^c \;.
\end{equation}
We expect  ${\cal S}_1$ to be a marginal operator (only) up to linear order
in $H$.

Now, one can apply conformal perturbation theory to compute the correction
to the three-point functions of three currents $J^a=i\partial X^a$, 
$\overline J^a=i\partial \overline X^a$. It turns out that 
there are also non-vanishing correlators like 
$\langle J^a J^b \overline J^c\rangle$, i.e. the currents are not 
holomorphic respectively anti-holomorphic. 
However, one can 
define new fields $\mathcal J^a$ and $\overline{\mathcal J}{}^a$ 
\begin{eqnarray}
  \label{def_cur_04}
\begin{split}
  {\cal J}^a (z,\overline z) & = J^a(z)-
  {\textstyle \frac{1}{2}}\hspace{0.5pt}H^a{}_{bc} \, J^b(z) \, 
  {X}^c_R(\overline z)  \;, \\[1mm]
  \overline{\cal J}^a (z,\overline z) & = \overline J{}^a(\overline z)-
  {\textstyle \frac{1}{2}}\hspace{0.5pt}H^a{}_{bc} \, X_L^b( z) \,
  \overline J{}^c(\overline z)  \;
\end{split}
\end{eqnarray}
so that the three current correlators take the CFT form
\begin{eqnarray}
  \label{cur_cor_19}
\begin{split}
 \bigl\langle {\cal J}^a(z_1,\overline z_1)\, {\cal J}^b(z_2,\overline z_2)\, {\cal J}^c(z_3,\overline z_3) \bigr\rangle  
 &=  -i\:\frac{{\alpha'}^2}{8}\,H^{abc}\:
   \frac{1}{ z_{12}\, z_{23}\, z_{13}} \;, \\
  \bigl\langle \overline{\cal J}{}^a(z_1,\overline z_1)\, \overline{\cal J}{}^b(z_2,\overline z_2)\,
  \overline {\cal J}{}^c(z_3,\overline z_3) \bigr\rangle 
  &= +i\:\frac{{\alpha'}^2}{8}\,H^{abc}\:
   \frac{1}{ \overline z_{12}\, \overline z_{23}\, \overline z_{13}} \;.
\end{split}
\end{eqnarray}
The necessity of this redefinition  can already be understood from the 
two-dimensional equation of motion
$\partial\overline\partial X^a={1\over 2} H^a{}_{bc} \partial X^b
\overline\partial X^c$. Therefore, already at linear order the 
coordinate fields have to be adjusted to be consistent with a CFT description.
However, the deformation is still marginal and nothing starts
to run.

Writing the new currents as derivatives of corrected coordinates 
${\cal X}^a$, after three integrations the three-point function
of these coordinates can be computed as
\begin{eqnarray}
  \label{tpf_01aa}
  \bigl\langle {\cal X}^a(z_1,\overline z_1)\, {\cal X}^b(z_2,\overline z_2)\,   {\cal X}^{c}(z_3,\overline z_3) \bigr\rangle^{H}   
= \theta^{abc} 
   \Bigl[ {\cal L} \Bigl( {\textstyle \frac{z_{12}}{z_{13}} } \Bigr)
   -  {\cal L} \Bigl( {\textstyle \frac{\overline z_{12}}{\overline z_{13}} } \Bigr) \Bigr]
\end{eqnarray}
with $\theta^{abc}=\frac{{\alpha'}^2}{12}\: H^{abc}$ and 
\begin{equation}
{\cal L}(z)=
L(z) 
  + L \Bigl( 1-{1\over x} \Bigr) 
  + L \Bigl( {1\over 1-x} \Bigr)\, ,
\end{equation}
where the Rogers dilogarithm is defined as
\begin{equation}
  L(z)={\rm Li}_2(z) + \frac{1}{ 2} \log (z) \log(1-z)\; .
\end{equation}
It satisfies the so-called fundamental  identity  $L(z)+L(1-z)=L(1)$.
The three-point function \eqref{tpf_01aa} should be considered
as the closed string genera\-lization of the second term in
\eqref{twopointa}. However, there two essential differences:
\begin{itemize}
\item{For the closed string it is the three- and not the two-point function
which is corrected.}
\item{For the closed  string the Rogers dilogarithm  gives rise to
a non-trivial world-sheet dependence,
whereas for the open string only the essentially constant 
step-function appeared.}
\end{itemize} 
One can also compute the correction to the two-point function of two
coordinates. It reads 
\begin{equation}
   \delta_2 \bigl\langle X^a(z_1,\overline z_1) \, X^b(z_2,\overline z_2)  \bigr\rangle =
   \frac{{\alpha'}^2}{8} H^a{}_{pq}\, H^{bpq} \, \log |z_1-z_2|^2\;  \log\epsilon \;,
\end{equation}
where $\epsilon$ is a cut-off. 
Therefore, we explicitly see that the perturbation ${\cal S}_1$ ceases 
to be marginal at second order in the flux. The theory is no longer 
conformally invariant and starts to run according to the 
renormalization group flow equation for the inverse world-sheet metric $G^{ab}$,
which is  of the form 
\begin{equation}
  \mu\: \frac{\partial\hspace{0.5pt} G^{ab}}{ \partial \mu} =-\frac{\alpha'}{4} 
   H^a{}_{pq}\, H^{bpq}  \; .
\end{equation}
This  precisely agrees with equation \eqref{targeteom} for constant space-time metric,  $H$-flux and dilaton.

Up to linear order in the flux we can  write the energy-momentum tensor
as
\begin{equation}
  \label{def_emt_01}
  {\cal T}(z) = \frac{1}{\alpha'}\: \delta_{ab}:\! {\cal J}^a {\cal J}^b\!:\! (z) \;, \hspace{40pt}
  \overline{\cal T}(\overline z) = \frac{1}{\alpha'}\: \delta_{ab} :\! \overline{\cal J}{}^a\overline{\cal J}{}^b\!:\! (\overline z) \;.
\end{equation}
They give rise to  two copies of the Virasoro algebra with central charge
$c=3$ and ${\cal J}^a(\overline{\cal J}{}^a)$ is indeed a (anti-)chiral 
primary field with $h=1(\overline h=1)$.

The aim is to carry out a similar computation as for the open
string case, i.e. to evaluate string scattering amplitudes
for vertex operators and to see whether there is any sign
of a new space-time noncommutative/nonassociative product.
Recall that in the free theory the tachyon vertex operator
is a primary field  of conformal dimension $(h,\overline h)=(\frac{\alpha'}{4}\,
p^2,\frac{\alpha'}{4}\, p^2)$, and
in covariant quantization of the bosonic string
physical states are given by primary fields of conformal
dimension $(h,\overline h)=(1,1)$. 
The natural definition of the tachyon vertex operator 
for the perturbed theory is
\begin{equation}
  \label{def_vo7023}
  {\cal V}(z,\overline z) 
= \, :\!\exp \bigl( i\hspace{0.5pt} p \cdot ({\cal X}_L +
  {\cal X}_R) \bigr) \!: \;.
\end{equation}
One can compute
\begin{eqnarray}
\begin{split}
   {\cal T}(z_1)\, {\cal V}(z_2,\overline z_2) 
   &=
   \frac{1}{(z_1-z_2)^2} \, \frac{\alpha' p\cdot p}{4} \,   {\cal V}(z_2,\overline z_2)  +\\[-0.1cm]
    &\qquad\qquad\qquad\quad\frac{1}{z_1-z_2} \, \partial {\cal V}(z_2,\overline z_2) 
   +{\rm reg.} \;,
\end{split}
\end{eqnarray}
and analogously for the anti-holomorphic part.
This means that the vertex operator \eqref{def_vo7023} is  primary and 
has conformal dimension $(h,\overline h)=(\frac{\alpha'}{4}\,
p^2,\frac{\alpha'}{4}\, p^2)=(1,1)$. It is therefore a physical quantum
state of the  deformed theory.

\section{T-duality, R-flux and tachyon amplitudes}

Even though in the framework of the Buscher rules, applying three
T-dualities on the $H$-flux background is questionable, 
on the level of the CFT, T-duality corresponds to a simple
asymmetric transformation of the world-sheet theory.
It is just a reflection of the right-moving coordinates. 
Since our  corrected fields ${\cal X}^a(z,\overline z)$  still admit
a split into a holomorphic and an anti-holomorphic piece,
we define T-duality on the  world-sheet action 
along the direction $\mathcal X^a$ as 
\begin{equation}
  \begin{array}{c}
  {\cal X}_L^a(z) \\[1mm]
  {\cal  X}_R^a(\overline z)   
  \end{array}
  \qquad
  \xrightarrow{\;\mbox{\scriptsize T-duality}\;}
  \qquad
  \begin{array}{c}
  +{\cal X}_L^a(z)\;, \\[1mm]
  -{\cal  X}_R^a(\overline z) \;.
  \end{array}
\end{equation}
Under a T-duality in all three directions,  momentum modes in the
$H$-flux background are mapped to winding modes in the $R$-flux
background. We are now interested in momentum modes in the
R-flux background  which are related via T-duality to
winding modes in the $H$-flux background. 
Therefore, the  three-point function in the $R$-flux background should
read
\begin{eqnarray}
  \label{tpf_01}
  \bigl\langle {\cal X}^a(z_1,\overline z_1)\, {\cal X}^b(z_2,\overline z_2)\,   {\cal X}^{c}(z_3,\overline z_3) \bigr\rangle^{R}   
= \theta^{abc} 
   \Bigl[ {\cal L} \Bigl( {\textstyle \frac{z_{12}}{z_{13}} } \Bigr)
   +  {\cal L} \Bigl( {\textstyle \frac{\overline z_{12}}{\overline z_{13}} }
   \Bigr) \Bigr]\, ,
\end{eqnarray}
which just has a different relative sign between the holomorphic
and anti-holomorphic part. Here, we have the relation
$\theta^{abc}=\frac{{\alpha'}^2}{12}\: R^{abc}$.

For the correlator of three tachyon vertex operators one obtains
\begin{eqnarray}
  \label{threetachyonb}
\begin{split}
   \bigl\langle \,{\cal V}_1 \,{\cal V}_2 \,{\cal V}_3 \,\bigr\rangle^{H/R}
   &=\frac{\delta(p_1+p_2+p_3)}{\vert z_{12}\,z_{13}\,z_{23}\vert^2}\times\\
     & \qquad \exp\Bigl[ -i\hspace{0.5pt}\theta^{abc}\, p_{1,a} p_{2,b} p_{3,c}  \bigl[
  {\cal L}   \bigl( {\textstyle \frac{z_{12}}{z_{13}} }\bigr) \mp {\cal L}
   \bigl({\textstyle \frac{\overline z_{12}}{ \overline z_{13}}}\bigr)\bigr] \Bigr]_{\theta} \,,
\end{split}
\end{eqnarray}
where $[\ldots]_{\theta}$ indicates that the result is valid only up to linear order in $\theta$.
The full string scattering amplitude of the integrated tachyon vertex
operators then becomes
\begin{eqnarray}
  \label{threetacyon}
\begin{split}
  \bigl\langle\, \mathcal T_1\: \mathcal T_2\:\mathcal  T_3\, \bigr\rangle^{H/R}
  &= \int \prod_{i=1}^3  d^2 z_i\, \delta^{(2)}(z_i-z_i^0) \, \delta(p_1+p_2+p_3) \times \\
 & \hspace{40pt}
 \exp\Bigl[ -i\hspace{0.5pt}\theta^{abc}\, p_{1,a} p_{2,b} p_{3,c}  \bigl[
  {\cal L}   \bigl( {\textstyle \frac{z_{12}}{z_{13}} }\bigr) \mp {\cal L}
   \bigl({\textstyle \frac{\overline z_{12}}{ \overline z_{13}}}\bigr)\bigr] \Bigr]_{\theta}.
\end{split}
\end{eqnarray}

Let us now study the behavior of \eqref{threetachyonb}  under
permutations of the vertex operators. 
Before applying momentum conservation, 
the three-tachyon amplitude for a permutation $\sigma$ of
the vertex operators can be computed 
using the  relation $L(z)+L(1-z)=L(1)$. 
With $\epsilon=-1$ for the $H$-flux and $\epsilon=+1$ for the $R$-flux, 
one finds
\begin{eqnarray}
  \label{phasethreeperm}
\begin{split}
&\bigl\langle \, {\cal V}_{\sigma(1)}   {\cal V}_{\sigma(2)}  {\cal V}_{\sigma(3)}  \bigr\rangle^{H/R}=\\
&\qquad\qquad\qquad \exp\Bigl[ \,i\left({\textstyle \frac{1+\epsilon}{ 2}}\right)  \eta_\sigma\,  \pi^2\,  \theta^{abc}\, p_{1,a} 
  \,p_{2,b} \,p_{3,c} \Bigr]
  \bigl\langle {\cal V}_1\,  {\cal V}_2\,  {\cal V}_3  \bigr\rangle^{H/R} \;,
\end{split}
\end{eqnarray}
where in addition $\eta_{\sigma}=1$ for an odd permutation and 
$\eta_{\sigma}=0$ for an even one. 
One observes that for  $H$-flux the phase is always trivial while 
for  $R$-flux  a non-trivial phase may appear.
Recall that our analysis is only reliable up to linear order in $\theta^{abc}$.

Note that it is  non-trivial  that this phase is independent
of the world-sheet coordinates, which can be traced back  to
the form of the fundamental identity of $L(z)$. 
For this reason, it can be thought of as a property of the underlying target 
space.
Indeed, the phase in \eqref{phasethreeperm} can be recovered from 
a new three-product on the space of  functions
$V_{p_n}(x)=\exp( i\, p_n \cdot x )$  which is defined as 
\begin{eqnarray}
  \label{threebracketexp}
\begin{split}
   &V_{p_1}(x)\,\tri\, V_{p_2}(x)\, \tri\, V_{p_3}(x)\stackrel{\rm def}{=} \\
   &\qquad\qquad\qquad\qquad \exp\Bigl( -i \,{\textstyle \frac{\pi^2}{2}}\, \theta^{abc}\,
   p_{1,a}\, p_{2,b}\, p_{3,c} \Bigr) V_{p_1+p_2+p_3}(x)\; .
\end{split}
\end{eqnarray}
However, in CFT correlation functions operators
are understood to be radially ordered and so changing the order
of operators should not change the form of the amplitude.
This is known as crossing symmetry.
In the case of the $R$-flux background, this
is reconciled by applying momentum conservation leading to
\begin{equation}
  p_{1,a} \,p_{2,b}\, p_{3,c}\,\theta^{abc} = 0
  \hspace{40pt}{\rm for}\hspace{40pt} p_3=-p_1-p_2 \;.
\end{equation}
Therefore, scattering amplitudes of three tachyons do not receive
any corrections at linear order in $\theta$  both for the $H$- and
 $R$-flux.  

The three-product \eqref{threebracketexp} can be generalized to 
 more generic functions as
\begin{eqnarray}
\label{threebracketcon}
\begin{split}
  & f_1(x)\,\tri\, f_2(x)\, \tri\, f_3(x) \stackrel{\rm def}{=} \\
&\qquad\qquad\qquad\exp\Bigl(
   {\textstyle {\pi^2\over 2}}\, \theta^{abc}\,
      \partial^{x_1}_{a}\,\partial^{x_2}_{b}\,\partial^{x_3}_{c} \Bigr)\, f_1(x_1)\, f_2(x_2)\,
   f_3(x_3)\Bigr|_{x} \;,
\end{split}
\end{eqnarray}
where we used the notation $(\ )\vert_x=(\ )\vert_{x_1=x_2=x_3=x}$.
This is to be compared with the $\star$-product \eqref{Nbracketcon}
and can be thought of as a possible closed string generalization
of the open string noncommutative structure.
Note that \eqref{threebracketcon} is precisely the three-product anticipated in
an analysis of the $SU(2)$ WZW model\cite{arXiv:1010.1263}.
Indeed, the three-bracket for the coordinates $x^a$ can then be re-derived  as
the completely antisymmetrized sum of three-products
\begin{equation}
\label{antisymtripcon}
   \bigl [x^a,x^b,x^c \bigr]=\sum_{\sigma\in P^3} {\rm sign}(\sigma) \;  
     x^{\sigma(a)}\, \tri\,  x^{\sigma(b)}\, \tri\,  x^{\sigma(c)} =
     3\pi^2\, \theta^{abc}\; ,
\end{equation}
where $P^3$ denotes the permutation group of three elements.
For the WZW model\cite{arXiv:1010.1263}, this three-bracket was defined
as the Jacobi-identity \eqref{jacobiid} of the coordinates, which can
only be non-zero if the space is noncommutative and nonassociative
(see also the similar paper by L\"ust\cite{arXiv:1010.1361}).

This result generalizes to the $N$-tachyon amplitude, where
the  relative phase can be described by the following deformed product
\begin{eqnarray}
\begin{split}
   &f_1(x)\, \tri_N\,  f_2(x)\, \tri_N \ldots \tri_N\,  f_N(x) \stackrel{\rm def}{=} \\
   &\exp\left[ {\textstyle {\pi^2\over 2}} \theta^{abc}\!\!\!\!\! \sum_{1\le i< j < k\le N}
     \!\!\!\!  \, 
      \partial^{x_i}_{a}\,\partial^{x_j}_{b} \partial^{x_k}_{c} \right]\, 
   f_1(x_1)\, f_2(x_2)\ldots
   f_N(x_N)\Bigr|_{x} \;, 
\end{split}
\end{eqnarray}
which is the closed string generalization of the open string
noncommutative product \eqref{Nbracketcon}.
The phase becomes trivial after taking momentum conservation into account
or equivalently 
\begin{eqnarray}
  \int d^n x f_1(x)\, \tri_N\,  f_2(x)\, \tri_N \ldots \tri_N\,  f_N(x) 
   = \int d^n x f_1(x)\,f_2(x)\, \ldots \,  f_N(x)\, .\quad 
\end{eqnarray}

\section{Comments}

We have used conformal perturbation theory  to analyze the
bosonic string moving 
in an $H$- respectively $R$-flux background, at least up to linear
order in the flux. In the $R$-flux case, the application of  T-duality
to the string scattering amplitudes of tachyon vertex operators
revealed  a non-trivial three-product structure which was
visible, however, only prior to implementing momentum conversation.
At first sight this might be puzzling, but it actually makes
sense. If we had found a non-vanishing phase factor for
a closed string scattering amplitude, it would have been  in clear
conflict with crossing symmetry of CFT amplitudes. 
Another way of saying this is: The deformation of space-time as
implied by a non-vanishing three-bracket for the coordinates is
consistent with the structure of two-dimensional CFT.
In view of the fact that asymmetric CFTs are known to not
admit  a usual geometric target-space interpretation, this is
an interesting observation.

In the original paper \cite{arXiv:1106.0316}, we also computed the complete
four-tachyon scattering amplitude. It could be written in  
the  $SL(2,\mathbb C)$ invariant and expli\-citly crossing symmetric
form  
\begin{eqnarray}
\label{fourtachyons}
 &&\langle {\cal T}_1\, {\cal T}_2\, {\cal T}_3\, {\cal T}_4\rangle^{H/R}=\\[0.1cm]
 &&\hspace{0pt}
 \int d^2 X\:  {\exp\Bigl[-i\hspace{0.5pt} \theta^{abc}\,   p^1_a\, p^2_b\, p^3_c\,  
     \bigl[ (-{3\over 2}L(1) + {\cal L}(X)) \mp (-{3\over 2}L(1) + {\cal L}(\overline X)) \bigr] \Bigr]_{\theta} \over
     \vert X\vert^{2-2a}\; \vert 1-X\vert^{2-2c} } \;,\nonumber
\end{eqnarray}
where $X$ denotes the cross-ratio 
$X=\frac{(z_1-z_2)(z_3-z_4)}{(z_1-z_3)(z_2-z_4)}$ and 
\begin{equation}
  \label{some_def_6329}
     a={\alpha'\over 4}(p_1+p_4)^2-1 \;, \hspace{10pt}
     b={\alpha'\over 4}(p_1+p_3)^2-1 \;, \hspace{10pt}
     c={\alpha'\over 4}(p_1+p_2)^2-1\;.
\end{equation}
The three corresponding Mandelstam variables read $u=-(p_1+p_4)^2$,
$t=-(p_1+p_3)^2$ and $s=-(p_1+p_2)^2$, and the on-shell external tachyons
satisfy $\alpha' p_i^2=4$ so that $a+b+c=1$. 
Equation \eqref{fourtachyons} is  the fluxed version  of the 
Virasoro-Shapiro amplitude. 
The analysis of the pole-structure revealed that the spectrum
of states changes at linear order in the flux. For instance, 
some of the former massless modes, like the graviton,  seem
to become massive and some  even tachyonic.
This is conceptually consistent with the observation that
the vertex operator of  the ``graviton'' 
\begin{equation}
  \label{gravvert}
  {\cal V}_G(z,\overline z) 
  =  G_{ab}   :\! {\cal J}^a \, \overline{\cal J}{}^b  \exp \bigl(
  i\hspace{0.5pt} p \cdot {\cal X}\bigr) 
\end{equation}
is generically not any longer a primary field of conformal dimension
one and therefore not a physical state.

So far, only  correlation functions involving tachyons were analyzed.
In contrast to the open string case, one expects that vertex operators of the 
form \eqref{gravvert} will also contain new contractions between
the $H/R$-flux and the polarizations. Recall that for the open
string they were absent,  as  $\theta^{ab}$ was
multiplied by a piecewise constant function (see eq.\eqref{twopointa}).
This is clearly not true for the Rogers dilogarithm.

Let me close with another observation.
Ignoring for the moment that the graviton $G$ and the analogously defined
two-form $B$
vertex operators \eqref{gravvert} are not physical, the computation
of the $\langle GGB\rangle,\langle BGB\rangle$ scattering
amplitudes contain  $SL(2,\mathbb Z)$ invariant contributions
of the schematic form
\begin{eqnarray}
\begin{split}
   &H{\rm -flux}:\qquad \langle GGB\rangle\simeq \theta^{abc} \,
  G_{am} p_3^m\,  G_{b}{}^{n} B_{cn}+\ldots \; , \\[0.1cm]
   &R{\rm -flux}:\qquad\,  \langle BGB\rangle\simeq \theta^{abc} \, B_{am}\, p^m_3
  \, G_{b}{}^n\,  B_{cn} +\ldots \,  .
\end{split}
\end{eqnarray}
For $H$-flux the second amplitude is vanishing and for $R$-flux
the first.  In the first case, such a term arises from the
term $H^{abc} H_{abc}$ in the effective action with $H=dB$.  
The second amplitude, however, rather suggests that 
the effective action contains a term $R^{abc} R_{abc}$ with
$R^{abc}= B^{am}\, \partial_m  B^{bc} +\ldots$.
This is very similar to the form of the non-geometric $R$-flux as it appears
for instance in double field theory \cite{arXiv:1106.4015}.
In this context, $B$ is usually denoted as $\beta$ and is rather 
a bi-vector than a two-form.     
Thus, the question arises whether one can formulate something like
an effective action for these non-geometric fluxes where
the structures presented in this talk might play an important role.

I deeply regret that all these exciting questions cannot be 
discussed with Max anymore.

\section*{Acknowledgments}
I would like thank Andreas Deser, Dieter L\"ust, Erik Plauschinn and Felix
Rennecke for collaboration and many discussions 
on the material  presented in this talk. I am also very grateful
to the organizers of the ``Memorial Conference for Maximilian Kreuzer''
to give me the opportunity to  speak.


\end{document}